\documentstyle[aps,preprint]{revtex}
\tightenlines
\begin{document}
\preprint{TRI-PP-98-42}
\date{May 12, 1999}

\title{Reply to the comment on 'Validity of certain soft photon amplitudes'}
\author{Mark Welsh {\footnote{email: markw@retrologic.com} } 
and Harold W. Fearing {\footnote{email: fearing@triumf.ca}}}
\vskip 20pt
\address{TRIUMF, 4004 Wesbrook Mall, Vancouver, British Columbia,Canada V6T 
2A3}
\nopagebreak
\maketitle
\begin{abstract}
We respond to the accompanying Comment on our paper, 'Validity of certain soft 
photon amplitudes'. While we hope the discussion here clarifies the issues, we 
have found nothing which leads to a change in the original conclusions of our 
paper. 
\end{abstract}  

In Ref.\ {\cite{I}}, hereafter referred to as I, we discussed some problems
which arise with generalized soft photon approximations (SPA). Such generalized
SPA's result from the infinite number of ways one can choose variables to
describe the elastic amplitude. Such choices give identical elastic amplitudes
when momenta satisfying $p_1+p_2=p_3+p_4$ are used. However when the variables
are evaluated using the radiative momenta, $p_1+p_2=p_3+p_4+k$, as is done in
the derivation of a SPA, the results differ. Thus one obtains generalized
SPA's, which however differ only by terms $O(k)$.  Our paper was motivated in
part by Ref.\ {\cite{Liou1}}, which applied the so-called TuTts and TsTts
generalized SPA's to proton-proton bremsstrahlung. However the work of I was
intended to be much more general and not just a comment on Ref.\
{\cite{Liou1}}.

In the accompanying Comment {\cite{Lioucom}} Liou, {\it et al.} raised two
objections to I. While we find these objections invalid, and stand by the
conclusions of the original paper, we wish to discuss them, as they reemphasize
some of the ambiguities which appear in these generalized SPA's.

In I we originally discussed two kinds of problems. The first dealt with phase
space and arises because for certain choices of variables the elastic
amplitude, evaluated using the radiative momenta, is required outside the
region which is physically measurable. This is the case for the TsTts
amplitude, if evaluated for proton-proton bremsstrahlung as described in detail
by Liou, {\it et al.} [Ref.\ {\cite{Liou1}}, p. 375, 2nd column]. We learned
however in the course of discussion of this Comment that the evaluation was
actually not done as described but instead by using a prescription (TETAS)
which in effect moved the variables back into the physical region as
necessary. This bypasses, albeit in an ad hoc fashion, the first difficulty
raised in I.

The first objection raised by Liou, {\it et al.} deals with the second problem
discussed in I, which had to do with the proper (anti)symmetrization of
amplitudes involving identical particles. We will discuss here only the spin
zero case for which the amplitudes must be symmetric. The generalization to
spin one-half and antisymmetric amplitudes is obvious as here spin is truly an
'inessential complication.' The original TuTts amplitude, reproduced in Eq.
(26) of I, but originally from {\cite{Liou3}}, was not properly symmetrized.
Symmetrizing in $p_3 \leftrightarrow p_4$ leads to Eq. (27) of I, which however
cannot obviously be expressed in terms of properly symmetrized elastic
amplitudes, which is necessary for a SPA expressed in terms of measurable non
radiative quantities. The problem arises with the ad hoc terms $\Delta_i$ which
are added to force gauge invariance. In the discussion following Eq.  (30) we
derived sufficient conditions on the $\Delta_i$'s to ensure that the SPA
amplitude, valid through $O(k^0)$, was gauge invariant, could be expressed in
terms of symmetrized elastic amplitudes, and had the proper analyticity
properties, namely the $O(k^0)$ terms do not have pieces of $O(k/k)$, all as
required by the general principles of SPA. We then gave four examples for the
$\Delta_i$ chosen from the infinite set of possibilities. Liou, {\it et al.}
also recognized this symmetrization problem and in a paper \cite{Liou2}
subsequent to their original one obtained an amplitude corresponding to one of
the $\Delta_i$ examples we had given.

In their Comment Liou, {\it et al.} work through a lot of algebra, but the
essence of their claim is that their particular choice of $\Delta_i$ is the
only correct one because the others have $k/k$ type structures at $O(k)$,
i.~e. terms of $O(k^2/k)$. Such structures do exist as they claim, but in our
view they are irrelevant for a discussion of the validity of a SPA {\it as a
soft photon approximation.} As is well known, gauge invariance provides only
one condition and is sufficient only to fix the $O(k^0)$ terms. Thus SPA's are
valid only through $O(k^0)$. There will be many terms at $O(k)$, including some
with structure $O(k^2/k)$ coming from higher order expansions of the external
radiation graphs, which are just not determined in a SPA and thus are not
relevant to SPA discussions. This principle, that what happens at $O(k)$ is
immaterial, is actually acknowledged by Liou, {\it et al.} in their Comment,
below Eq. (6), in reference to their second objection.

On the other hand if one wants to make a model dependent choice among the
infinite set of SPA's to find one that fits data, as was done in Ref.\
{\cite{Liou1}}, but not in I, then such external criteria, while having nothing
to do with SPA as such, would be relevant and clearly in the absence of other
information one would choose the $\Delta_i$ which was in some sense
'smoothest'.

The second objection of the Comment deals with the size of the error incurred
in using a TuTts amplitude which is not properly symmetrized. The authors of
the Comment attribute to us a much more general statement than we intended or
actually worked out in I. Specifically in I we compared Eq. (27), which was the
TuTts amplitude properly symmetrized, with Eq. (28) which was the prescription
we understood was used in Ref.\ {\cite{Liou1}} to obtain numerical results
using a TuTts amplitude which did not have the proper symmetry properties. All
of our remarks concerning the relative size of corrections refer to this
specific comparison.  Subsequently Liou, {\it et al.} \cite{Liou3} derived a
properly symmetrized amplitude to be used in their numerical calculations, thus
making a prescription unnecessary, and our choices of $\Delta_i$ lead to
similar amplitudes. Thus in our opinion this comparison has been superseded and
is not particularly important. However it does illustrate yet a further
ambiguity in such SPA's, and so for that reason is perhaps worth
discussing. The problem arises because of the multiple ways one can define a
properly symmetrized elastic amplitude. We define one such amplitude in
Eq. (29) of I for one case, which happens to be simultaneously symmetric in
$p_3 \leftrightarrow p_4$ and $u \leftrightarrow t$. For the other amplitudes
however there is a choice, because of the freedom to define the elastic
variables in different ways. For example we could define $A^{\prime
S}(u_{23},t_{13}) = A^\prime (u_{23},t_{13})+ A^\prime (t_{24},u_{14})$ which
is symmetric in $p_3 \leftrightarrow p_4$ but not in $u \leftrightarrow t$ and
which might be appropriate if the elastic amplitude comes from a diagrammatic
calculation easily expressed in terms of the $p_i$. Alternatively we could take
$A^{\prime S}(u_{23},t_{13}) = A^\prime (u_{23},t_{13})+ A^\prime
(t_{13},u_{23})$ which is symmetric in $u \leftrightarrow t$ but not in $p_3
\leftrightarrow p_4$ and which might be more appropriate for an elastic
amplitude taken from numerical data given as a function of energy and angle.
For the elastic amplitude evaluated for the nonradiative momenta
$p_1+p_2=p_3+p_4$ these are of course identical, but they are formally
different functions of the momenta so they differ, by terms of $O(k)$, when
evaluated using the radiative momenta $p_1+p_2=p_3+p_4+k$. Thus when we compare
Eq. (27) and (28) using the first of these, as we did in I, the difference is
$O(k/k)$ as was stated, while if we use the second, as apparently is being done
by Liou, {\it et al.} in the context of their second objection, the difference
is $O(k)$. Clearly this difference can arise only because Eq. (28) is a
prescription which may not be a 'real' SPA (cf. discussion following Eq. (29)
in I) since we know that 'real' SPA's can differ only at $O(k)$.

In summary, in our view neither of the objections raised by Liou, {\it et al.}
in their Comment lead to any changes in the conclusions of I. They do however
serve to reemphasize the main point of I, namely that there are a lot of
ambiguities related to the use of these generalized SPA's. One is always faced
with a dilemma. If $k$ is small, the $O(k)$ terms are small and all SPA's are
the same and essentially model independent. On the other hand if $k$ is larger,
as is the case for all modern measurements of proton-proton bremsstrahlung,
then the $O(k)$ terms make a difference and one must make a whole set of model
dependent choices which basically have nothing to do with SPA's {\it per se},
but which can affect the quality of fit to the data and the predictive power,
or lack thereof, of the model.

This work was supported in part by the Natural Sciences and Engineering
Research Council of Canada.

\end{document}